PAPER • OPEN ACCESS

# The baryon loading effect on relativistic astrophysical jet transport in the interstellar medium

PAPER

# The baryon loading effect on relativistic astrophysical jet transport in the interstellar medium

W P Yao[1], B Qiao[1,2,3,4,6,7], Z Xu[1], H Zhang[1,4], Z H Zhao[1], H X Chang[1], C T Zhou[1,4], S P Zhu[3,5] and X T He[1,2,3]

[1] Center for Applied Physics and Technology, HEDPS, and State Key Laboratory of Nuclear Physics and Technology, School of Physics, Peking University, Beijing 100871, People's Republic of China
[2] Collaborative Innovation Center of IFSA (CICIFSA), Shanghai Jiao Tong University, Shanghai 200240, People's Republic of China
[3] Institute of Applied Physics and Computational Mathematics, Beijing 100094, People's Republic of China
[4] Center for Advanced Material Diagnostic Technology, Shenzhen Technology University, Shenzhen 518118, People's Republic of China
[5] Graduate School of China Academy of Engineering Physics, PO Box 2101, Beijing 100088, People's Republic of China
[6] Collaborative Innovation Center of Extreme Optics, Shanxi University, Taiyuan, Shanxi 030006, People's Republic of China
[7] Author to whom any correspondence should be addressed.

**E-mail:** bqiao@pku.edu.cn

## Abstract

The composition of the astrophysical relativistic jets remains uncertain. By kinetic particle-in-cell simulations, we show that the baryon component in the jet, or the so-called baryon loading effect (BLE), heavily affects relativistic jets transport dynamics in the interstellar medium. On the one hand, with the BLE, relativistic jets can transport in a much longer distance, because jet electrons draw a significant amount of energy from jet baryons via the Buneman-induced electrostatic waves and the Weibel-mediated collisionless shock; on the other hand, the jet electron phase space distribution may transform from a bottom-wide-single-peak structure to a center-wide-multiple-peak one by increasing the BLE, which largely influences the observed jet morphology. Implications for related astrophysical studies are also discussed.

## 1. Introduction

Astrophysical jets exist widely in the Universe [1–3]. Among them, relativistic jets are often connected with the observed exciting high-energy astrophysical phenomena, such as gamma-ray burst [4–6], active galactic nucleus (AGNs) [7–9], tidal disruption events [10, 11] and compact binary stars [12, 13]. Recently, the *Fermi* Gamma-Ray Space Telescope has detected a typical AGN [14], in which the jet bulk Lorentz factor is about 10 $\sim$ 20. However, despite the growing interest in relativistic astrophysical jets, our understanding of them is still limited. Two main reasons are as follows.

On the one hand, because these high-energy astrophysical phenomena are so far away from Earth, *in situ* probes are unavailable. As a result, many fundamental parameters of relativistic jets are still under debate, especially the jet composition. For simplification purposes, purely leptonic jet components (electrons and positrons) are often assumed. However, recent studies [15, 16] have identified that both leptons and baryons are presented in the relativistic jets. In fact, the influence of the baryon components on the relativistic jet transport in interstellar medium (ISM) is uncertain yet.

On the other hand, the observed morphology of the relativistic jets during their transport in the ISM has been extensively investigated. Specifically speaking, various dichotomies have been given, such as Fanaroff–Riley (FR) type-I/type-II [17–20] and bottom-wide/center-wide [21–23], etc. It is generally thought that the macroscopic hydrodynamic instabilities, e.g. Rayleigh–Taylor [24] and Kelvin–Helmholtz [25, 26] instabilities (KHI), result in the formation of these different morphology. However, to give a more consistent physical explanation of the jet morphology, the microscopic particle dynamics need to be seriously considered [27], especially in regions such as the X-point of magnetic reconnection [28–31] and the discontinuity surface of collisionless shock [32]. From this point of view, kinetic instabilities, with the spatial scale length at the order of the plasma skin depth, have been identified as playing important roles, e.g. the Buneman instability (BI) [33–37] and the Weibel instability (WBI) [38, 39].

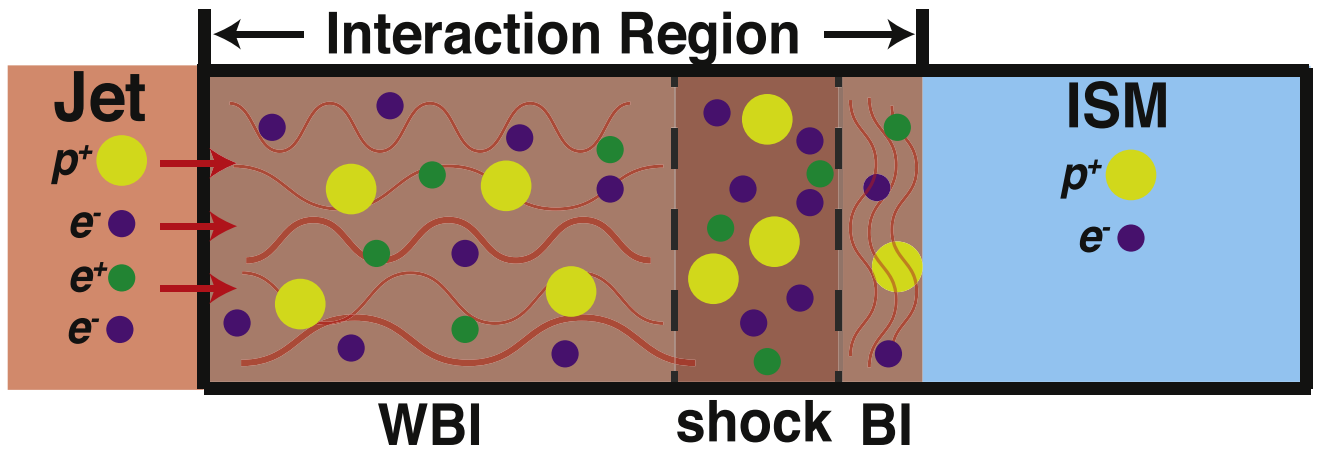

**Figure 1.** Schematic for 2D PIC simulation setup on modeling relativistic jet–ISM interaction. Just like [41, 42], the simulation box (black) is initially filled up with ISM (light blue), which composes of protons ($p^+$) and electrons ($e^-$), and relativistic jets (light red) continuously inject from the left, composed of a certain ratio of protons, electrons and positrons ($e^+$). In relativistic jet–ISM interaction region, the red stripes represent the density modulations caused by both BI and WBI. And the dashed darker region represent the collisionless shock when WBI saturates.

In this paper, the effect of the baryon component, or the so-called baryon loading effect (BLE) [40], on the relativistic astrophysical jet transport dynamics in the ISM has been investigated theoretically and by fully kinetic particle-in-cell (PIC) simulations with a novel particle injection module. From a series of PIC simulations of relativistic jets transport in the ISM with various initial parameters including densities, drifting velocities, compositions, and distributions, we show clearly that with the BLE, the jet can transport in a much longer distance, because the jet electrons draw a significant amount of energy from the jet baryons via the Buneman-induced ES waves and the Weibel-mediated shock waves. Moreover, by increasing the BLE, the jet phase space distribution transforms from a bottom-wide-single-peak structure to a center-wide-multiple-peak one. As a wider spread in the transverse phase space distribution indicates a wider jet radius, the BLE can largely influence the observed jet morphology.

The paper is organized as follows. Section 2 introduces the numerical simulation setup. In section 3, representative simulation results demonstrate the BLE on both the relativistic jet transport dynamics and their morphology. To explain, kinetic instabilities and momentum phase space distributions are carefully analyzed in section 4. In section 5, parametric studies systematically show the robustness of our findings. Section 6 gives the conclusions and discussions.

## 2. Simulation setup

To self-consistently study the relativistic jet–ISM interaction from the microscopic particle dynamics, a series of two-dimensional (2D) PIC simulations are performed. Similar to the simulation model in previous studies of jet-ambient interaction [41, 42], the relativistic jet is injected from the left into the simulation box, which is filled up with ISM. As figure 1 shows, the simulation box has a length of $L_x = 5000\lambda_e$ and a width of $L_y = 50\lambda_e$, where $\lambda_e = c/\omega_{pe}$ is the electron skin depth. Here, $c$ is the light speed, $\omega_{pe} = (n_m q_e^2/\varepsilon_0 m_e)^{1/2}$ is the electron plasma frequency ($n_m$ is the ISM electron number density, $\varepsilon_0$, $m_e$ and $q_e$ are the vacuum permittivity, electron mass and electron charge, respectively). The simulation box is divided into $25000 \times 250$ cells along $x$ and $y$, respectively, with the Debye length $\lambda_{\mathrm{De}} = (\varepsilon_0 k T_{m0}/n_m q_e^2)^{1/2}$ resolved, where $T_{m0} = 1$keV is the initial ISM plasma temperature and $k$ is the Boltzmann constant. In order to clearly demonstrate how the jet evolves during its transport, the simulation time is kept rather long ($t_{\mathrm{sim}} = 5000\omega_{pe}^{-1}$). The ISM is composed of electrons and protons, each specie is represented by 8 macro-particles per cell. We use a reduced proton mass $m_p/m_e = 64$ due to the limitation of the computational resources, while it manages to ensure that the scale separation between electron and proton dynamics can be modeled in fully kinetic simulations [43]. The ISM species follow the Maxwell distribution $f_s \propto \exp(m_s c^2/T_s)(s = p, e)$, where the initial electron and proton temperatures are both 1 keV.

A novel injection module has been implemented in the code to generate a jet flowing into the simulation box from the left boundary at $x = 0$. Here the jet is charge-neutralized without the requirement of adding an additional field damping layer at the injection boundary, which is different from injecting an electron beam into plasmas [44]. Also, our injection module is different from the 'reflection method' commonly adopted in collisionless shock studies, which take advantage of the reflect boundary condition [39, 45]. Relativistic jets with various initial parameters including densities, drifting velocities, compositions, and distributions can be loaded from our injection module. Furthermore, in order to suppress the numerical Cherenkov noise [46] caused by the relativistic jet (with drifting velocity close to $c$), the 3rd-order B-Spline shape function, 5th-order weighting, and 6th-order finite different scheme for solving Maxwell's equations have been adopted. The jet is composed of electrons ($e^-$), positrons ($e^+$) and baryons (loading with protons ($p^+$)), which satisfies $n_{e^+} + n_{p^+} = n_{e^-} = n_j$ ($n_j$ represents the jet density). The ratio of the relativistic jet density $n_j$ to that of the ISM $n_m$ is chosen to be $\alpha = n_j/n_m = 1.1$ here, and the jet keeps injecting into the ISM continuously till the simulation ends. All species of the jet follow a drifted Maxwell–Jüttner distribution [47, 48]

$$f_s(\boldsymbol{p}) \propto \exp\left[-\frac{\gamma_d}{T_s}(\sqrt{1+\boldsymbol{p}^2} - V_d p_x)\right](s = p^+, e^+, e^-),$$

where the temperature $T_s = 1$ keV is taken. $\boldsymbol{p}$ is the momentum and $V_d$ is the drift velocity along $x$-direction. The corresponding Lorentz factor is $\gamma_d = (1 - (V_d/c)^2)^{-1/2} = 15$. The Sobol method [49] is used to load relativistic particles at each step.

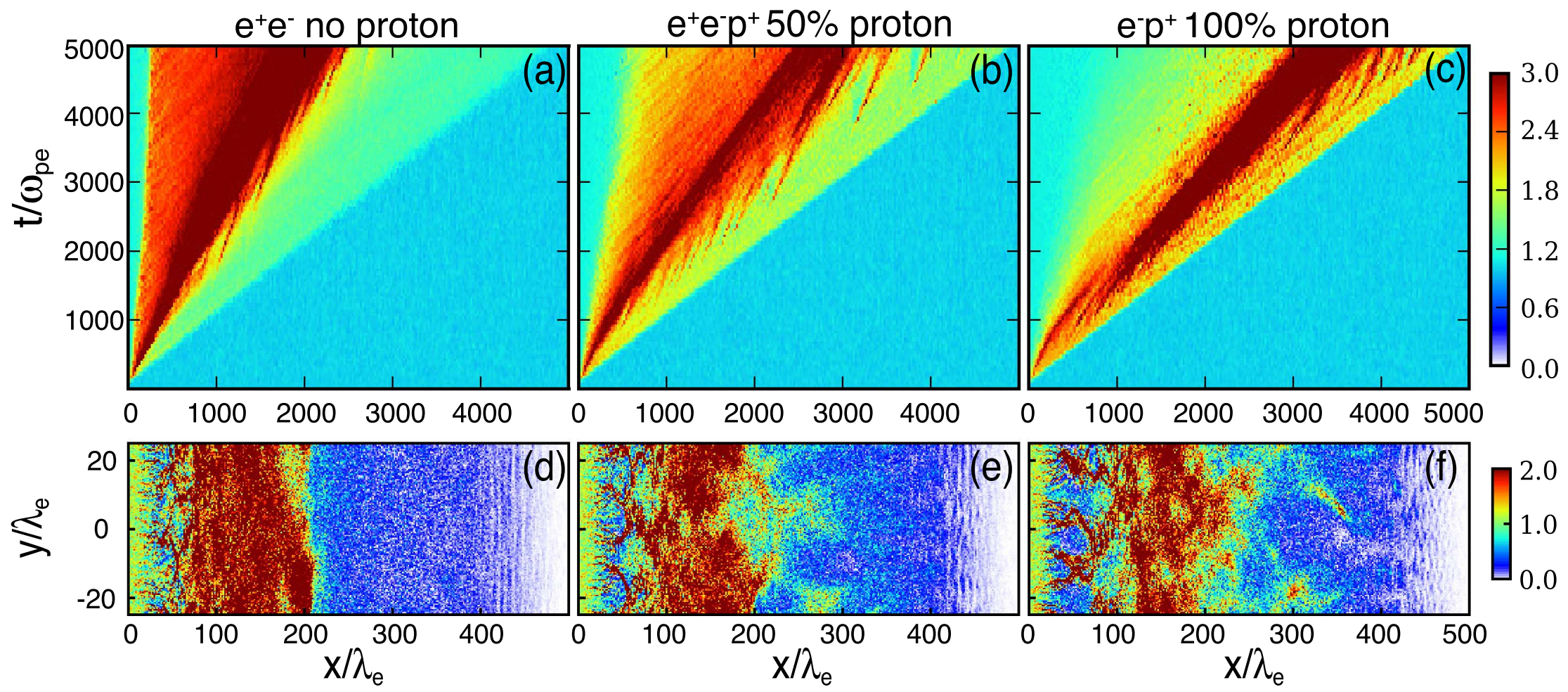

**Figure 2.** The first row is the time evolution of jet and ISM electron density (averaged transversely) for (a) $e^{\pm}$, (b) $e^-e^+p^+$ and (c) $e^-p^+$ case, respectively. The colorbar represents $n_j + n_m$ (normed by $n_m$). The second row is 2D distribution of jet electron density at $t = 500\omega_{pe}^{-1}$ for corresponding cases and the colorbar represents $n_j$ (also normed by $n_m$).

In the following sections, three typical simulation cases are demonstrated:

(i) $e^{\pm}$ case, the jet is composed of electrons and positrons only, i.e. the leptonic jets;

(ii) $e^-e^+p^+$ case, the jet is composed of electrons, positrons and protons with densities $n_{p^+} = n_{e^+} = 0.5n_{e^-}$, i.e. the lepton/baryon mixture jets;

(iii) $e^-p^+$ case, the jet is composed of electrons and protons only, i.e. the baryonic jets.

## 3. The BLE on relativistic jet transport dynamics

Certainly, one of the most concerned issues is how far the relativistic jet can stably transport in the ISM. Figure 2 shows a global picture of the jet transport dynamics for the above cases with/without the BLE. From the time evolution of electron densities in figures 2(a)–(c), when the ratio of the baryon component increases, the jet can transport in much longer distances. Specifically, without the BLE in $e^{\pm}$ case, the leptonic jet only transports for a distance of $x = 2100\lambda_e$ (see also in figure 5(a)); while adding 50% baryons in $e^-e^+p^+$ case, the jet transport distance is extended to about $x = 2700\lambda_e$ (see also in figure 5(b)). Moreover, for a purely baryonic jet in $e^-p^+$ case, the jet transports to a much longer distance of nearly $x = 3500\lambda_e$ (see also in figure 5(c)). From 2D distributions of jet electron density in figures 2(d)–(f) at $t = 500\omega_{pe}^{-1}$, when the BLE increases, we can see that the density modulations of the jet become more violent. Specifically, at the jet bottom region ($0 < x < 100\lambda_e$), transverse filamentation clearly develops, especially for the $e^-p^+$ case. By contrast, at the jet head region ($400\lambda_e < x < 500\lambda_e$), pronounced longitudinal modulations can been seen.

These density modulations are caused by kinetic instabilities. From 2D electromagnetic (EM) field distributions in figures 3(a)–(c) around $0 < x < 400\lambda_e$ at $t = 500\omega_{pe}^{-1}$, transverse EM modes occur for all cases with the spatial scales close to the ion skin depth $\lambda_i = c/\omega_{pi}$, where the WBI dominates and lead to the jet transverse density modulations [41, 42, 50, 51]. When the BLE plays role, $B_z$ becomes stronger and the WBI dominated region (WDR) significantly expands, corresponding to the extension of jet transport distance. Moreover, from 2D electrostatic (ES) field distributions in figures 3(d)–(f) around $400\lambda_e < x < 500\lambda_e$ at the same time, the ES waves can be seen in all cases with the wave length in electron skin depth scale ($\lambda_e = c/\omega_{pe}$), where the BI dominates and leads to jet longitudinal density modulations [52–54]. In essence, this longitudinal electric field and density modulation are caused by the ion density gradient at the leading edge of the jet, together with the different mobilities of electrons and ions induced by BLE [55]. To clearly see the BLE on the BI-induced ES wave, the integrals of the ES fields ($\phi = \int E_x d_x$) are shown in figures 3(g)–(i) at $380\lambda_e < x < 480\lambda_e$. With the increase of the BLE, the $\Delta\phi$ is also increased, indicating that when electrons transport through this region, they gain more energy from the baryons (see also in figures 5(d)–(f)). However, in $e^{\pm}$ case, the jet electrons and positrons cannot draw energy from each other. Instead, they gradually lose energy to the ISM. Thus, the leptonic jet transport in the ISM is a pure stopping process, resulting in a much limited transport distance. Note that during the relativistic jet–ISM interaction, the BI-induced ES mode is coexisting with the WBI-induced EM mode [39, 56, 57].

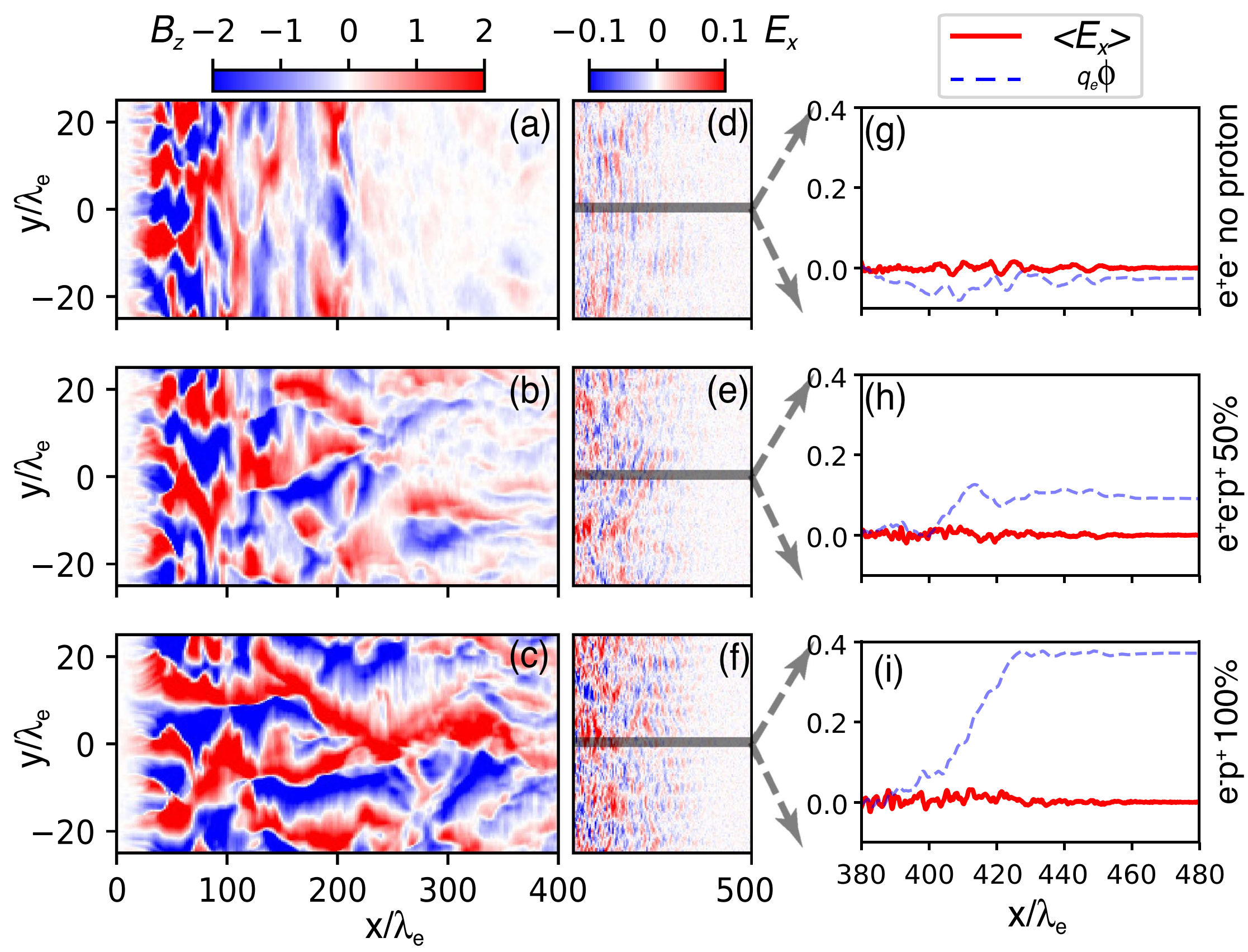

**Figure 3.** EM fields distribution at $t = 500\omega_{pe}^{-1}$. The left two columns are time-averaged $B_z$ ($0 < x < 400\lambda_e$, normed by $B_0 = m_e\omega_{pe}/q_e$) and $E_x$ ($400\lambda_e < x < 500\lambda_e$, normed by $E_0 = B_0 c$) for ((a), (d)) $e^{\pm}$, ((b), (e)) $e^-e^+p^+$ and ((c), (f)) $e^-p^+$ case, respectively. The right column is the axial distribution (averaged transversely) of $E_x$ and its integral $\phi = \int E_x d_x$ for corresponding cases.

From another perspective, we also see that the WBI-mediated collisionless shocks form and propagate in the ISM from figures 2(a)–(c). By tracing the slope of the piled-up electron density surface of $n_j + n_m$, the shock velocity can be estimated as $v_{\mathrm{sh},ee} \approx 0.42c$, $v_{\mathrm{sh},eep} \approx 0.54c$, and $v_{\mathrm{sh},ep} \approx 0.70c$, respectively. Assuming the upstream electron temperature of all cases are the same ($T_e = 1$ keV), the corresponding ion sound speed will be $C_s = \sqrt{kT_e/m_p} \approx 0.0055c$. Thus, their Mach numbers $M = v_{\mathrm{sh}}/C_s$ can be calculated as $M_{ee} \approx 76.02$, $M_{eep} \approx 97.74$ and $M_{ep} \approx 126.70$, which are significantly enhanced with the increase of the BLE. In fact, compared with leptonic ones, jets with baryon components provide more free energy for the WBI when injecting with the same drift speed, because the initial energy of each specie is $\varepsilon_{s0} = \gamma_d m_s V_d^2$ ($s = p^+, e^+, e^-$). As a result, with the increase of the BLE, electrons gain more energy from the baryons, leading to stronger collisionless shocks.

Additionally, to show the BLE on electron acceleration during the relativistic jet–ISM interaction, energy evolution and spectra are given in figure 4. Specifically, figure 4(a) shows the energy exchange of leptons and EM fields in $e^{\pm}$ and $e^-p^+$ case. As the jet continuously injecting into the ISM, in $e^-p^+$ case, more free energy are provided with baryons to the WBI. When the WBI saturates and collisonless shock forms, more energy is converted into magnetic field energy (see green lines) and then drawn by electrons (see red lines), compared with that in $e^{\pm}$ case. Similarly, the BI also gets more free energy and is converted into more electric field energy (see blue lines). While in figure 4(b), the energy spectra of jet electrons at $t = 500\omega_{pe}^{-1}$ and $t = 5000\omega_{pe}^{-1}$ are shown. In $e^{\pm}$ case, few electron acceleration is realized, where the maximum electron energy (normalized by $m_e c^2$ ) increases only from 90 to 150 (dashed arrow) during the transport. However, for $e^-p^+$ case, the maximum electron energy increased from about 100 to 450 (horizontal solid arrow). Moreover, the particle number of the high-energy electrons also substantially increases (vertical solid arrows). Note that electrons will keep gaining energy until magnetic field energy reach the sub-equipartition level [58].

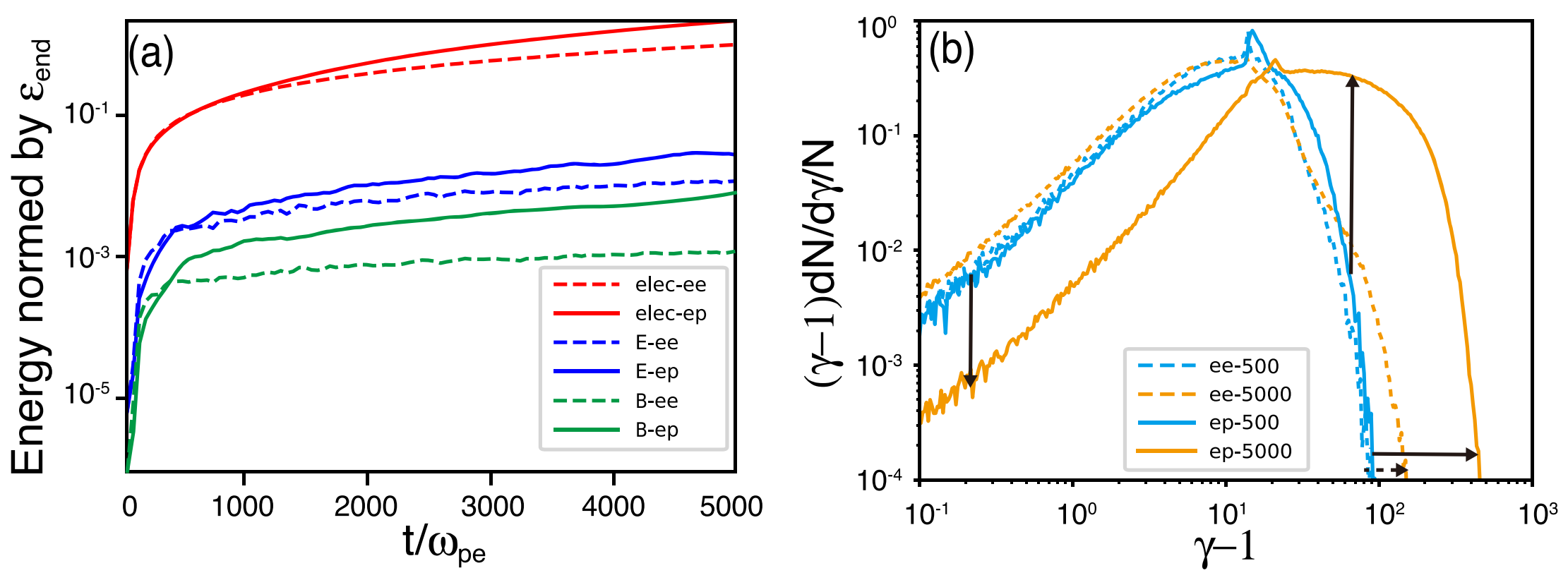

**Figure 4.** In (a) the energy evolution of two cases ($e^{\pm}$, dashed lines and $e^-p^+$, solid lines) are compared. Electron energy (red lines), electric field energy (blue lines) and magnetic field energy (green lines) are plotted for both cases (normed by the final electron energy of case $e^-p^+$ $\varepsilon_{\rm end}$). In (b) the jet electron energy spectrum at $t = 500\omega_{pe}^{-1}$ (cyan lines) and $t = 5000\omega_{pe}^{-1}$ (orange lines) are also plotted for both cases. Arrows (also dashed for $e^{\pm}$ and solid for $e^-p^+$) represent the change of the spectra.

## 4. The BLE on relativistic jet morphology

According to the morphology of relativistic astrophysical jets, they have been sorted into typical dichotomies, as mentioned in section 1. Generally speaking, the FR-I jet tends to decelerate in the ISM and is comparatively weak, while the FR-II jet is more powerful and spatially extended. Besides, the most luminous part of the bottom-wide jet lies at the bottom, while that of the center-wide one lies at the center. Owing to the lack of *in situ* probes, the exact composition of these relativistic jets remains unknown. By kinetic PIC simulations, our findings can provide several implications for jet compositions. For example, from the transport dynamics perspective in section 3, our results imply that FR-I jets should be dominated by leptons, while FR-II ones by baryons.

Similarly, the jet electron momentum phase space distributions also help interpreting the jet composition, as they largely determine the jet morphology. Although a 3D "global jet" model [26, 51] is required to properly simulate the jet morphology, here in our 2D simulation, a wide spread in the transverse phase space distribution (or, a large $p_y$) indicates wide jet radius. Specifically speaking, in $e^{\pm}$ case, without the BLE, jet electrons lose forward momentum during the transport (figure 5(d)), and the WDR remains at the jet bottom region ($0 < x < 2000\lambda_e$), where the electrons are heated and scattered via the WBI (figure 5(g)), forming a bottom-wide structure. However, in $e^-e^+p^+$ case, with a certain BLE, jet electrons gain momentums in both directions from jet baryons during the transport (figures 5(e) and (h)), which moves the WDR to the jet body region ($2000\lambda_e < x < 4000\lambda_e$). Moreover, in $e^-p^+$ case, jet electrons gain much more momentums during the transport both longitudinally and transversely (figures 5(f) and (i)), and the WDR extends completely to the jet body, forming a center-wide structure.

Note that at the leading edge of the jet with baryons, narrow longitudinal momentum peaks develop due to the BI-induced ES waves (figures 5(e) and (f)). Because those electrons already have a relativistic injection speed, they co-move with the ES wave towards upstream, leading to a long dephasing length and continuous acceleration. As a result, the jet phase space distribution transforms from a bottom-wide-single-peak structure to a center-wide-multiple-peak one, where we can interpret that the former should be dominated by leptons, while the latter by baryons. This also implies that the single-zone model [59] (commonly adopted in the jet emission calculation) may only suitable for leptonic jets. Moreover, apart from the diffusive shock acceleration [60–62], this acceleration at the leading edge of the relativistic jet may serve as another mechanism in generating the ultra-high-energy cosmic ray (UHECR) [58, 63].

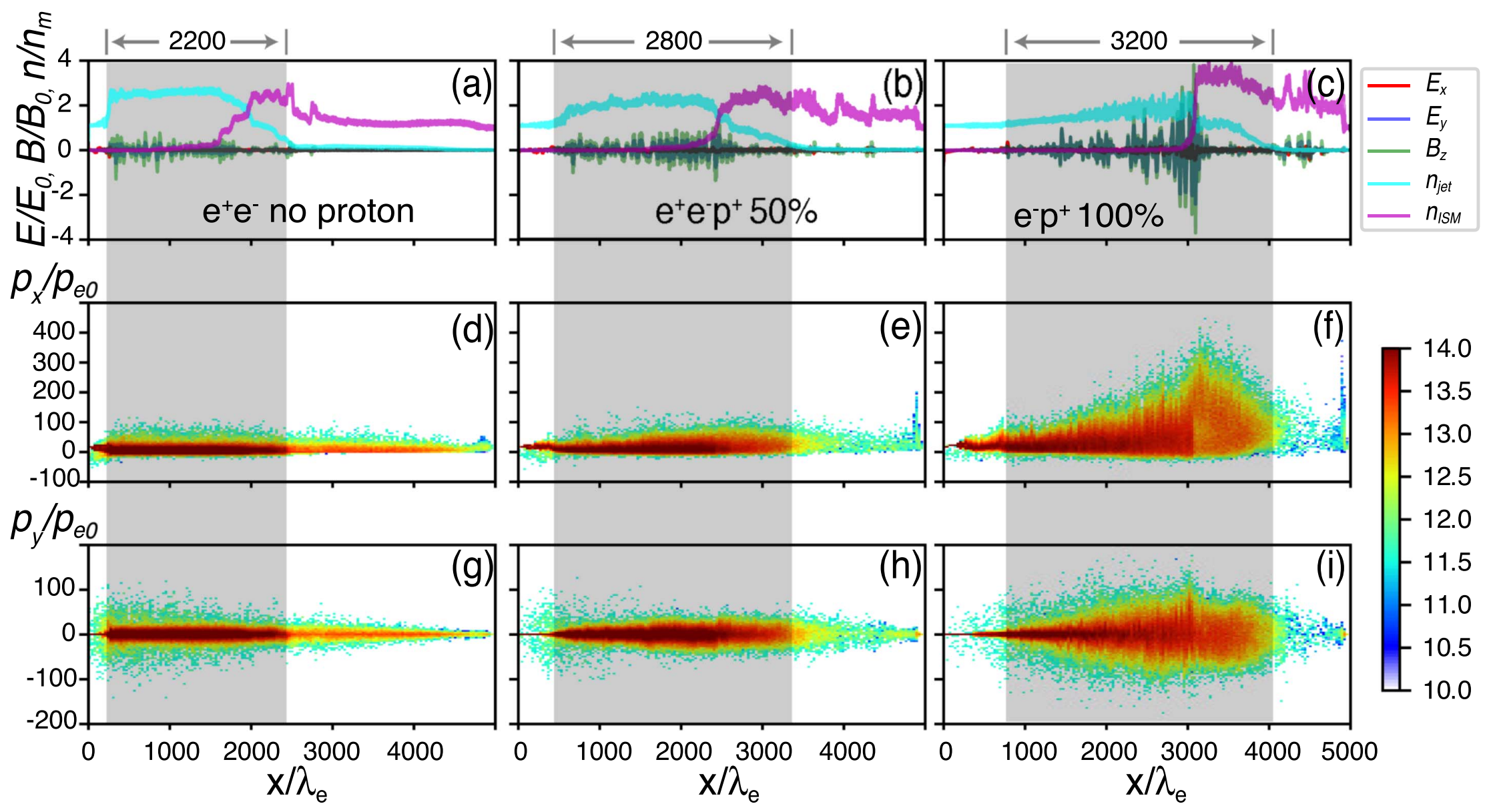

**Figure 5.** The first row is the axial-averaged distribution of electron densities and EM fields for (a) $e^{\pm}$, (b) $e^{-}e^{+}p^{+}$ and (c) $e^{-}p^{+}$ cases, respectively. The second row is the longitudinal momentum phase space ($x - p_x$, normed by $p_{e0} = m_e V_d$) of jet electrons for (d) $e^{\pm}$, (e) $e^{-}e^{+}p^{+}$ and (f) $e^{-}p^{+}$ cases, respectively. While the third row is the transverse momentum phase space ($x - p_y$, also normed by $p_{e0}$) of jet electrons for corresponding cases. All plots are the results at $t = 5000\omega_{pe}^{-1}$. Grey shaded areas are the WDR. The colorbar represents the macro-particle number in logarithm.

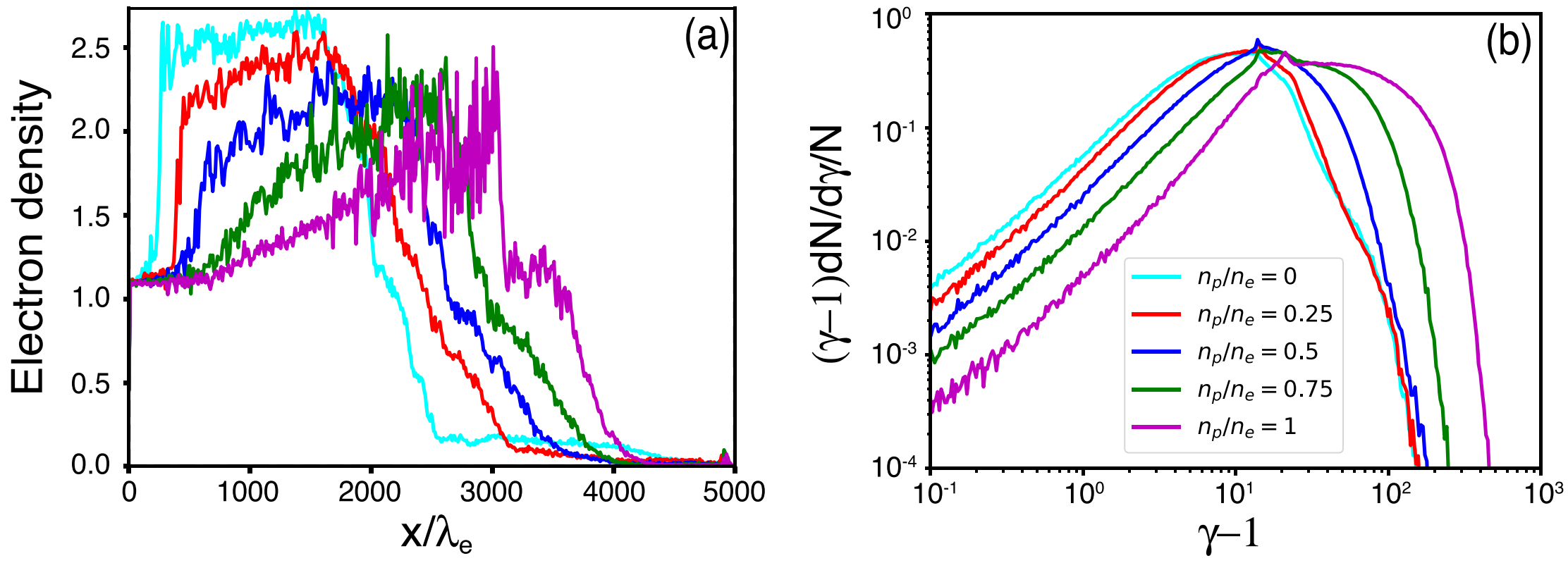

**Figure 6.** Parametric studies of the BLE on (a) jet transport distance (shown by jet electron density, normalized by its initial density $n_{e0}$) and (b) jet electron spectrum at $t = 5000\omega_{pe}^{-1}$. Five cases representing different amount of baryon component are given by solid lines with different colors (see the legend shared by both plots located in (b)). Note that in all the above cases, the density ratio of jet to ISM is the same, i.e. $\alpha = n_j/n_m = 1.1$, as mentioned in section 2.

## 5. Parameter Studies

Systematic parametric studies of the BLE on relativistic jets transport in the ISM are performed. Firstly, five cases with different baryon densities ($n_{p^+}/n_{e^-} = 0.00/0.25/0.50/0.75/1.00$) are presented in figure 6. Specifically speaking, with the increase of the baryon amount, the jet transport distance elongates from about $2100\lambda_e$ to about $3500\lambda_e$ in figure 6(a), and the maximum energy of the jet electrons increase from about 150 to nearly 450 in figure 6(b).

Secondly, the mass of the baryon component ($m_p/m_e$) influences the BLE. With larger $m_p/m_e$ (e.g. $m_p/m_e = 64/100/225/400$), the injected protons have more initial energy ($\varepsilon_0 = \gamma_d m_p V_d^2$) and the difference mobilities in each specie are larger, so that more energy can be drawn by jet electrons from kinetic instabilities, leading to stronger BLE. However, large $m_p/m_e$ increases the simulation time needed for the formation of collisionless shock, because it decreases the growth rate of the WBI. Thus, more computational resources are required.

Thirdly, the "hardness" of the jet (i.e. $\alpha = n_j/n_e$) also influences the BLE. With smaller $\alpha$ (e.g. $\alpha = 0.1/0.5$), the WBI growth rate is smaller (compared with the presented cases), and the simulation time needed for WBI saturation becomes longer. While a larger $\alpha$ (e.g. $\alpha = 5/10$) leads to stronger numerical noise, thus we have to adopt higher simulation spatial resolution. Because the BLE on relativistic jet–ISM interaction is all significant for the above cases ($m_p/m_e = 64 \sim 400, \alpha = 0.1 \sim 10$), only one of them ($m_p/m_e = 64, \alpha = 1.1$, with tolerable computational consumption and numerical noise) is presented here in sections 3 and 4.

## 6. Conclusions and discussions

In this paper, we have theoretically and numerically investigated the effect of the jet baryon component (or BLE) on the relativistic astrophysical jet transport dynamics in the ISM. Through a series of kinetic PIC simulations with various parameters, we find that, on the one hand, with BLE the jet transports in a much longer distance because the jet electrons/positrons draw a significant amount of energy from the jet baryons; on the other hand, the BLE largely influences the observed jet morphology, where the jet electron/positron phase space distribution transforms from a bottom-wide-single-peak structure to a center-wide-multiple-peak one by increasing the BLE.

These findings provide us several implications on astrophysical studies, especially the jet component, which is still under debate. To begin with, from the transport distance perspective, our findings imply that the FR-I jets should be dominated by leptons, while the FR-II ones by baryons. Moreover, from the jet phase space perspective, our findings imply that the bottom-wide-single-peak jets are likely to be composed of leptons, while the center-wide-multiple-peaks jets may be composed of baryons. Thus, the single-zone model may only be suitable for the leptonic jets. In addition, the continuous electron acceleration process at the jet leading edge may serve as another mechanism in generating UHECR. However, we have to admit that, to investigate the BLE on the global jet structure, not only the working surface (where WBI instability and collisionless shock appear), but also the lateral velocity shear (where KHI and Mushroom instability excite), should be taken into consideration, which requires a 3D global jet model [26, 51].

## Acknowledgments

This work is supported by Science Challenging Project, No. TZ2017005, the NSAF, Grant No. U1630246; the National Key Program of S&T Research and Development, Grant No. 2016YFA0401100; the National Natural Science Foundation of China, Grants No. 11575298, No. 91230205, No. 11575031, and No. 11175026; the National Basic Research 973 Projects No. 2013CBA01500 and No. 2013CB834100, and the National High-Tech 863 Project. BQ acknowledges the support from Thousand Young Talents Program of China. The computational resources are supported by the Special Program for Applied Research on Super Computation of the NSFC-Guangdong Joint Fund (the second phase). YWP would like to thank ZZQ for valuable discussions about the astrophysical backgrounds.

## ORCID iDs

W P Yao https://orcid.org/0000-0002-6017-9300
Z Xu https://orcid.org/0000-0003-0113-0010

## References

[1] Bridle A H and Perley R A 1984 *Annu. Rev. Astron. Astrophys.* **22** 319–58
[2] Vietri M *et al* 2003 *Astrophys. J.* **592** 378
[3] Fabian A C 2012 *Annu. Rev. Astron. Astrophys.* **50** 455
[4] Piran T 1999 *Phys. Rep.* **314** 575667
[5] Bykov A M and Treumann R A 2011 *Astron. Astrophys. Rev.* **19** 42
[6] Kumar P and Zhang B 2015 *Phys. Rep.* **561** 1
[7] de Gouveia Dal Pino E M 2005 *Adv. Space Res.* **35** 908–24
[8] Marscher A P *et al* 2008 *Nature* **452** 966
[9] Romero G E, Boettcher M, Marko S and Tavecchio F 2017 *Space Sci. Rev.* **207** 5
[10] Piran T, Sądowski A and Tchekhovskoy A 2015 *Mon. Not. R. Astron. Soc.* **453** 157
[11] Liu F K, Zhou Z Q, Cao R, Ho L C and Komossa S 2018 *Mon. Not. R. Astron. Soc.* **472** L99–103

[12] Migliari S and Fender R P 2006 *Mon. Not. R. Astron. Soc.* **366** 79
[13] Pavan L *et al* 2014 *Astron. Astrophys.* **562** A122
[14] Kynoch D, Landt H, Ward M J, Done C, Gardner E, Boisson C, Arrieta-Lobo M, Zech A, Steenbrugge K and Santaella M P 2017 *Mon. Not. R. Astron. Soc.* **475** 404–23
[15] Rees M J and Meszaros P 1992 *Mon. Not. R. Astron. Soc.* **258** 41P
[16] Lazzati D, Ghisellini G and Celotti A 1999 *Mon. Not. R. Astron. Soc.* **309** L13
[17] Fanaroff B L and Riley J M 1974 *Mon. Not. R. Astron. Soc.* **167** 31P
[18] Laing R A and Bridle A H 2002 *Mon. Not. R. Astron. Soc.* **336** 1161
[19] Best P N 2009 *Astron. Nachr.* **330** 184
[20] Hagino K, Stawarz Ł, Siemiginowska A, Cheung C C, Kozieł-Wierzbowska D, Szostek A, Madejski G, Harris D E, Simionescu A and Takahashi T 2015 *Astrophys. J.* **805** 101
[21] Fabian A C, Sanders J S, Ettori S, Taylor G B, Allen S W, Crawford C S, Iwasawa K, Johnstone R M and Ogle P M 2000 *Mon. Not. R. Astron. Soc.* **318** L65
[22] Reynolds C S *et al* 2015 *Astrophys. J.* **808** 154
[23] Wilson A S, Smith D A and Young A J 2006 *Astrophys. J.* **644** L9
[24] Pizzolato F and Soker N 2006 *Mon. Not. R. Astron. Soc.* **371** 1835
[25] Reynolds C S, McKernan B, Fabian A C, Stone J M and Vernaleo J C 2005 *Mon. Not. R. Astron. Soc.* **357** 242
[26] Nishikawa K-I *et al* 2016 *Astrophys. J.* **820** 94
[27] Marcowith A *et al* 2016 *Rep. Prog. Phys.* **79** 46901
[28] Zenitani S and Hoshino M 2001 *Astrophys. J.* **562** L63–6
[29] Zenitani S and Hoshino M 2007 *Astrophys. J.* **670** 702–26
[30] Sironi L and Spitkovsky A 2014 *Astrophys. J.* **783** L21
[31] Xu Z, Qiao B, Yao W P, Chang H X, Zhou C T, Zhu S P and He X T 2017 *Phys. Plasmas* **24** 92102
[32] Yao W P, Qiao B, Xu Z, Zhang H, Chang H X, Zhou C T, Zhu S P, Wang X G and He X T 2017 *Phys. Plasmas* **24** 82904
[33] Buneman O 1958 *Phys. Rev.* **112** 1504
[34] Buneman O 1958 *Phys. Rev. Lett.* **1** 8
[35] Ishihara O, Hirose A and Langdon A B 1980 *Phys. Rev. Lett.* **44** 1404
[36] Riquelme M A and Spitkovsky A 2011 *Astrophys. J.* **733** 63
[37] Matsumoto Y, Amano T and Hoshino M 2012 *Astrophys. J.* **755** 109
[38] Silva L O, Fonseca R A, Tonge J W, Mori W B, Dawson J M and Silva L O 2002 *Phys. Plasmas* **9** 2458
[39] Matsumoto Y, Amano T, Kato T N and Hoshino M 2017 *Phys. Rev. Lett.* **119** 105101
[40] Fiore M, Silva L O, Ren C, Tzoufras M A and Mori W B 2006 *Mon. Not. R. Astron. Soc.* **372** 1851
[41] Ardaneh K, Cai D, Nishikawa K-I and Lembège B 2015 *Astrophys. J.* **811** 57
[42] Ardaneh K, Cai D and Nishikawa K-I 2016 *Astrophys. J.* **827** 124
[43] Bret A and Dieckmann M E 2010 *Phys. Plasmas* **17** 032109
[44] Taguchi T, Antonsen T M and Mima K 2017 *J. Plasma Phys.* **83** 905830204
[45] Sironi L and Spitkovsky A 2011 *Astrophys. J.* **741** 39
[46] Vay J L and Godfrey B B 2014 *C. R. Mec.* **342** 610
[47] Melzani M, Winisdoerffer C, Walder R, Folini D, Favre J M, Krastanov S and Messmer P 2013 *Astron. Astrophys.* **558** A133
[48] Swisdak M 2013 *Phys. Plasmas* **20** 062110
[49] Zenitani S and Plasma Phys 2015 *Phys. Plasmas* **22** 042116
[50] Milosavljević M, Nakar E and Spitkovsky A 2006 *Astrophys. J.* **637** 765
[51] Nishikawa K-I, Niemiec J, Hardee P E, Medvedev M, Sol H, Mizuno Y, Zhang B, Pohl M, Oka M and Hartmann D H 2009 *Astrophys. J.* **698** L10
[52] Bret A, Firpo M C and Deutsch C 2005 *Phys. Rev. Lett.* **94** 1
[53] Bret A, Gremillet L, Bénisti D and Lefebvre E 2008 *Phys. Rev. Lett.* **100** 1
[54] Bret A, Gremillet L and Dieckmann M E 2010 *Phys. Plasmas* **17** 120501
[55] Choi E J, Min K, Nishikawa K-I and Choi C R 2014 *Phys. Plasmas* **21** 72905
[56] Amano T and Hoshino M 2009 *Astrophys. J.* **690** 244
[57] Matsumoto Y, Amano T and Hoshino M 2013 *Phys. Rev. Lett.* **111** 215003
[58] Sironi L, Spitkovsky A and Arons J 2013 *Astrophys. J.* **771** 54
[59] Gardner E and Done C 2018 *Mon. Not. R. Astron. Soc.* **473** 2639
[60] Axford W I, Leer E and Skadron G 1977 *Int. Cosmic Ray Conf.* vol 11 p 132
[61] Bell A R 1978 *Mon. Not. R. Astron. Soc.* **182** 147
[62] Blandford R and Ostriker J P 1978 *Astrophys. J. Lett.* **221** L29
[63] Iwamoto M, Amano T, Hoshino M and Matsumoto Y 2017 *Astrophys. J.* **840** 52